# Chapter 17

# Safety

*C. Adorisio, I. Bejar Alonso\*, J.C. Gascon, T. Otto and S. Roesler*

CERN, Geneva, Switzerland

## 17 Safety

### 17.1 Radiation to personnel

17.1.1 Design constraints

Design constraints for new or upgraded facilities should ensure that the exposure of persons working on CERN sites, the public, and the environment will remain below the specified dose limits [1] under normal as well as abnormal conditions of operation, and that the optimization principle is implemented [2, 3]. In particular, the following design constraints apply.

The design of components and equipment must be optimized such that installation, maintenance, repair, and dismantling work does not lead to an effective dose, e.g. as calculated with Monte Carlo simulations, exceeding 2 mSv per person and per intervention. The design is to be revised if the dose estimate exceeds this value for cooling times compatible with operational scenarios.

The annual effective dose to any member of a reference group outside of the CERN boundaries must not exceed 10 µSv. The estimate must include all exposure pathways and all contributing facilities.

The selection of construction material must consider activation properties to optimize dose to personnel and to minimize the production of radioactive waste. In order to guide the user a web-based code (ActiWiz) is available for CERN accelerators [4].

17.1.2 The As Low As Reasonably Achievable (ALARA) principle in the design of the Long Straight Sections

Proton–proton collisions in the LHC experiments produce a secondary radiation field that penetrates into the adjacent accelerator tunnels and can cause severe activation of beam-line elements. Consequently, in such areas the design of components and infrastructure has to be optimized to follow the as low as reasonably achievable (ALARA) principle. The optimization of the design and later intervention is an iterative process: dose equivalent maps per unit time of exposure (called dose rate maps below) of the concerned area(s) are compiled from measurements and/or simulations with Monte Carlo particle transport codes such as FLUKA [5, 6]. Based on these maps, the personal and collective doses of the intervention teams are calculated by using an intervention plan that then allows identification of and optimizing of critical work steps in order to reduce doses. If the latter involves a change in design or work scenario, then doses are re-evaluated by repeating the above steps.

---

\* Corresponding author: Isabel.BejarAlonso@cern.ch

### 17.1.3 The FLUKA Monte Carlo code for radiation protection studies

The use of general-purpose particle interaction and transport Monte Carlo codes is often the most accurate and efficient choice for assessing radiation protection quantities at accelerators. Due to the vast spread of such codes to all areas of particle physics and the associated extensive benchmarking with experimental data, modelling has reached an unprecedented level of accuracy. Furthermore, most codes allow the user to simulate all aspects of a high energy particle cascade in one and the same run: from the first interaction of a TeV particle to the transport and re-interactions (hadronic and electromagnetic) of the produced secondaries, to detailed nuclear fragmentation, the calculation of radioactive decays, and even the electromagnetic shower caused by the radiation from such decays.

FLUKA is a general-purpose particle interaction and transport code with roots in radiation protection studies at high energy accelerators [5, 6]. It therefore comprises all features needed in this area of application.

Detailed hadronic and nuclear interaction models cover the entire energy range of particle interactions at the LHC, from energies of thermal neutrons to interactions of 7 TeV protons. Moreover, the interface with DPMJET3 [7] also allows the simulation of minimum-bias proton–proton and heavy ion collisions at the experimental interaction points, which enormously facilitates calculations of stray radiation fields around LHC experiments.

FLUKA includes unique capabilities for studies of induced radioactivity, especially with regard to nuclide production, their decay and the transport of residual radiation. Particle cascades by prompt and residual radiation are simulated in parallel based on microscopic models for nuclide production and a solution of the Bateman equations [8] for activity build-up and radioactive decay. The decay radiation and its associated electromagnetic cascade are internally flagged as such in order to distinguish them from the prompt cascade. This allows the user to apply different transport thresholds and biasing options to residual and prompt radiation and to score both independently.

Particle fluence can be multiplied with energy-dependent conversion coefficients to effective dose or ambient dose equivalents [9] at scoring time. Prompt and residual dose equivalent can thus be computed in three-dimensional meshes, the latter for arbitrary user-defined irradiation and cooling profiles.

An integral part of the FLUKA code development is benchmarking of new features against experimental data. It includes both the comparison of predictions of individual models to measurement results (e.g. nuclide production cross-sections) as well as benchmarks for actual complex situations as, for example, arising during accelerator operation.

### 17.1.4 FLUKA simulations

Comprehensive dose rate maps were calculated with FLUKA for the part of Long Straight Section 1 (LSS1) that extends from the Target Absorber Secondary (TAS) to the separation dipole D1 (so-called 'inner triplet area'), including the inner triplet quadrupole magnets and the corrector package (CP) according to the latest HiLumi LHC layout [10]. The results may also serve as guideline values for dose planning at inner triplet regions of LSS5, due to the similar design of the two LSSs.

The simulations were limited to the high energy secondary radiation field arising from the p–p collisions as they dominate the activation in these areas. The influences due to losses of the beam directed towards the interaction point (IP) and beam–gas interactions are not considered. The FLUKA geometry shown below represents the tunnel on the righthand side of LSS1.

The studies were done for 7 + 7 TeV p–p collisions at two different average luminosity values of $5 \times 10^{34}$ cm$^{-2}$ s$^{-1}$ (for the so-called 'nominal scenario') and $7.5 \times 10^{34}$ cm$^{-2}$ s$^{-1}$ (for the so-called 'ultimate scenario') with a 295 µrad half-angle vertical crossing in IP1 using DPMJET-III as the event generator.

The irradiation profiles used are based on the operational scenarios reported in Table 17-1.

Table 17-1: LHC operational parameters. The second and third columns refer to the so-called 'nominal scenario', which will lead to a total integrated luminosity of 3060 fb$^{-1}$. The fourth and fifth columns refer to the so-called 'ultimate scenario'. This scenario will lead to a total integrated luminosity of 3910 fb$^{-1}$. For the total integrated luminosity 310 fb$^{-1}$ are taken into account as integrated in the operational period before LS3.

| Shutdown | Year of LHC operation | Levelled luminosity [cm$^{-2}$ s$^{-1}$] | Integrated luminosity [fb$^{-1}$] | Levelled luminosity [cm$^{-2}$ s$^{-1}$] | Integrated luminosity [fb$^{-1}$] |
|---|---|---|---|---|---|
| **LS3** | | | | | |
| | 2026 | $5.00 \times 10^{34}$ | 250 | $7.50 \times 10^{34}$ | 300 |
| | 2027 | $5.00 \times 10^{34}$ | 250 | $7.50 \times 10^{34}$ | 300 |
| | 2028 | $5.00 \times 10^{34}$ | 250 | $7.50 \times 10^{34}$ | 300 |
| **LS4** | | | | | |
| | 2030 | $5.00 \times 10^{34}$ | 250 | $7.50 \times 10^{34}$ | 300 |
| | 2031 | $5.00 \times 10^{34}$ | 250 | $7.50 \times 10^{34}$ | 300 |
| | 2032 | $5.00 \times 10^{34}$ | 250 | $7.50 \times 10^{34}$ | 300 |
| **LS5** | | | | | |
| | 2034 | $5.00 \times 10^{34}$ | 250 | $7.50 \times 10^{34}$ | 300 |
| | 2035 | $5.00 \times 10^{34}$ | 250 | $7.50 \times 10^{34}$ | 300 |
| | 2036 | $5.00 \times 10^{34}$ | 250 | $7.50 \times 10^{34}$ | 300 |
| **LS6** | | | | | |
| | 2038 | $5.00 \times 10^{34}$ | 250 | $7.50 \times 10^{34}$ | 300 |
| | 2039 | $5.00 \times 10^{34}$ | 250 | $7.50 \times 10^{34}$ | 300 |
| | 2040 | | | $7.50 \times 10^{34}$ | 300 |

For the inelastic pp cross-section the value of 85 mb is used, on the basis of the extrapolation [11]. Cycles of three years of continuous operation followed by a long showdown (LS) of a duration of one year were repeated until integrated luminosities of about 3000 fb$^{-1}$ and 4000 fb$^{-1}$ are reached for the nominal and ultimate parameters, respectively.

The studies use a FLUKA implementation of the High Luminosity LHC inner triplet region developed by the FLUKA team [12] according to the latest design, mechanical layout, and specifications, including a detailed model of the inner triplet region with new large-aperture Nb$_3$Sn magnets (150 mm coil aperture), field maps, corrector packages, and segmented tungsten inner absorbers. As mentioned above, the inner triplet regions at LSS1 and LSS5 are radiologically equivalent and symmetrical around the interaction point. Thus, the simulations were performed for the righthand side of LSS1 only.

### 17.1.5 Results

Three-dimensional residual dose rate maps have been calculated from around 18 m distance from the interaction point up to around 82 m distance (i.e. from TAS to D1), and for six different cooling times: 1 hour, 1 day, 1 week, 1 month, 4 months, and 1 year after the beam stop for both the nominal and the ultimate scenarios, i.e. when 3000 fb$^{-1}$ and 4000 fb$^{-1}$, respectively, have been reached.

In order to give examples for the available results, residual dose equivalent rates are reported in the following as two-dimensional maps for the one month cooling time as well as profile plots at four different cooling times (1 week, 1 month, 4 months, and 1 year). The values correspond to the average over 30 cm around the beam pipe height and between 40–50 cm distance from the outer surface of the cryostat (closest possible working distance).

Figure 17-1 shows the ambient dose equivalent rate maps in the inner triplet area after one month of cooling time and Figure 17-2 shows the residual dose rate profiles in the aisle at 40 cm distance from the cryostat at four different cooling times, when a total integrated luminosity of 3060 fb$^{-1}$ is reached.

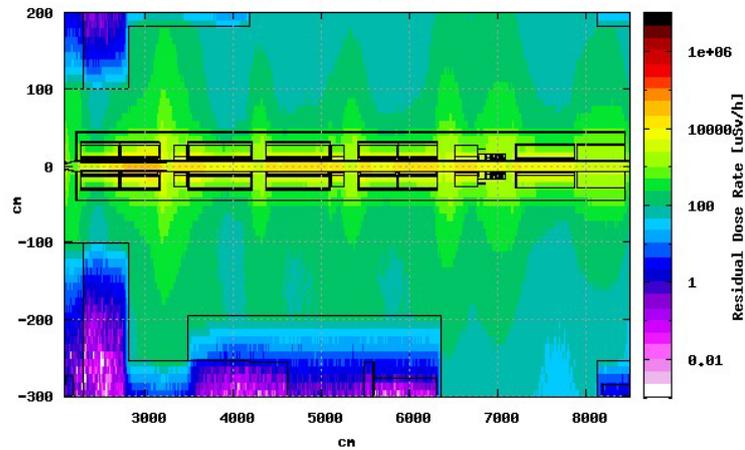

Figure 17-1: Residual dose equivalent rates in units of μSv/h around the inner triplet and D1 magnets when 3060 fb$^{-1}$ has been reached after one month of cooling time. Doses are shown for a horizontal section at the level of the beam lines. The origin of the coordinate frame is at the interaction point.

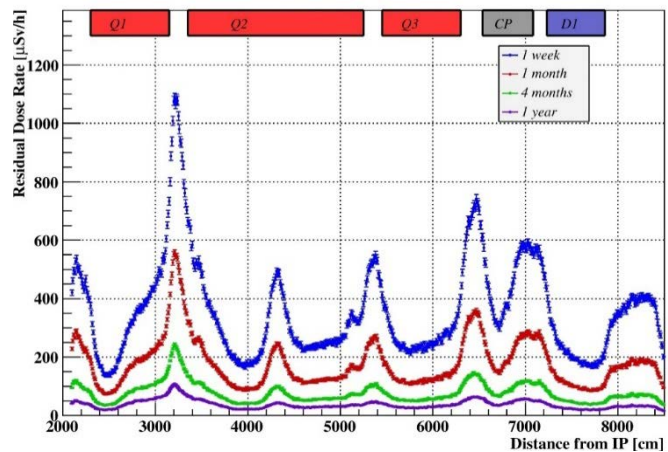

Figure 17-2: Ambient dose equivalent rate profiles for the nominal scenario in the aisle at 40 cm from the cryostat at four different cooling times: 1 week in blue, 1 month in red, 4 months in green and 1 year in violet.

Figure 17-3 shows the ambient dose equivalent rate maps in the inner triplet area after 1 month of cooling time and Figure 17-4 shows the residual dose rate profiles in the aisle at 40 cm distance from the cryostat at four different cooling times, for a total integrated luminosity of 3910 fb$^{-1}$.

The highest radiation levels will be found next to the magnet interconnections and in front of the TAS due to the 'self-shielding' provided elsewhere by the magnets themselves.

The plot in Figure 17-5 shows the residual dose rate profiles at 1 month cooling time along the inner triplets, for the ultimate scenario in pink and for the nominal scenario in light blue. The average ratio between the two profiles is 1.35. This average ratio varies with cooling time and it depends solely on the instantaneous luminosity in the case of short cooling times, while for longer cooling times the ratio also depends on the total integrated luminosity.

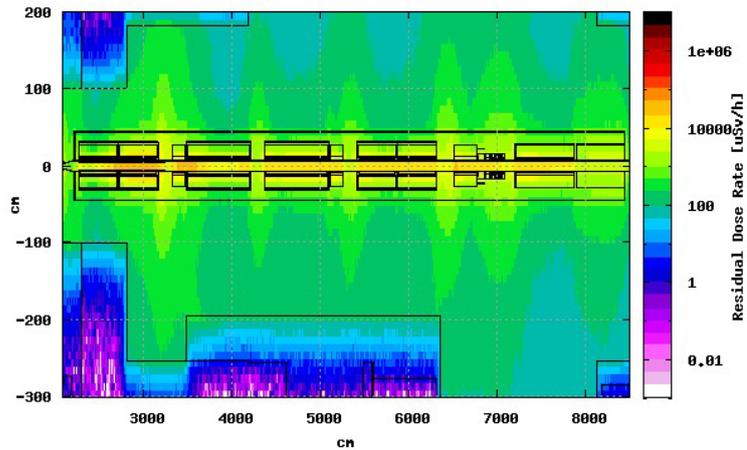

Figure 17-3: Residual dose equivalent rates in units of μSv/h around the inner triplet and D1 magnets when 3910 fb$^{-1}$ has been reached after one month of cooling time. Doses are shown for a horizontal section at the level of the beam lines. The origin of the coordinate frame is at the interaction point.

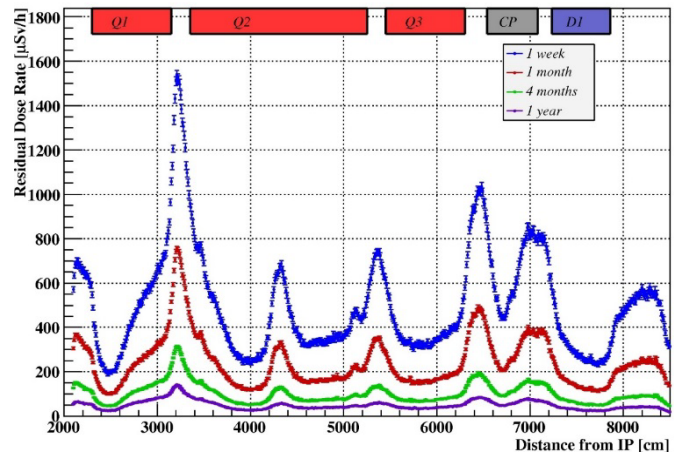

Figure 17-4: Ambient dose equivalent rate profiles for the ultimate scenario in the aisle at 40 cm from the cryostat at four different cooling times: 1 week in blue, 1 month in red, 4 months in green and 1 year in violet.

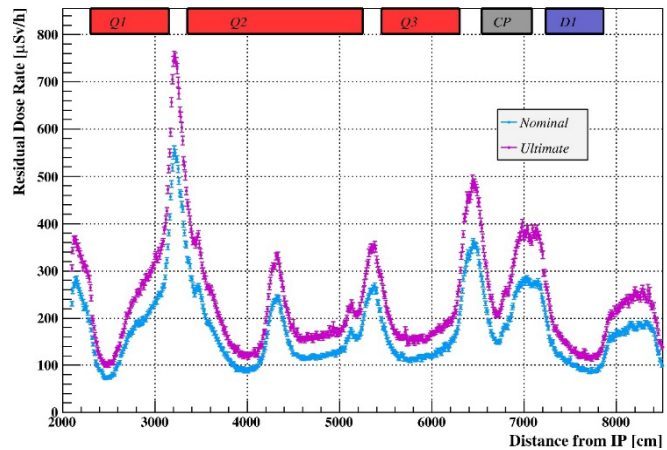

Figure 17-5: Ambient dose equivalent rate profiles along the aisle at 40 cm distance to the cryostat at one month cooling time. The profile in light blue is for the nominal scenario (about 3000 fb$^{-1}$ total integrated luminosity), the profile in pink is for the ultimate scenario (about 4000 fb$^{-1}$ total integrated luminosity).

17.1.6    Conclusions

Three-dimensional ambient dose equivalent rate maps were calculated with the FLUKA Monte Carlo code for the inner triplet region (from the TAS up to the D1 dipole magnet) for nominal and ultimate HiLumi LHC operational parameters and the latest upgrade layout. These predictions are now available and can be considered for the design of components. For example, the results will be used to assess the radiological impact, in terms of individual and collective dose, for typical interventions in the area and, thus, help in optimizing the design and the material choice as required by the ALARA principle. Further studies are also required to address airborne activation and the related environmental impact of the upgrade.

**17.2    General safety**

17.2.1    Implementation of safety in the HL-LHC

The HL-LHC presents a number of potential hazards that could pose serious safety risks. ̵ The consequences could include the loss of life, damage to HL-LHC systems and facilities, or damage to the environment. Consequently, all of the HL-LHC work packages (WPs) are classified as projects with major safety implications.

In practice the systems/processes of the HL-LHC WPs shall be submitted for a compulsory safety verification by the HSE Unit and shall only operate once the HSE Unit has granted safety clearance.

In the framework of the HL-LHC project, clearance will be given for the project phase, i.e. any development tests that will be performed on given systems or processes. After the HL-LHC project phase, a second clearance stage may become necessary once the system/process is ready for operation in the LHC.

17.2.2    Launch Safety Agreement (LSA)

For each HL-LHC work package a launch safety discussion shall take place with the participation of the WP members, the Project Safety Officer (PSO), and the correspondent from the safety unit (HSE-CO). After the launch safety discussion, the HSE-CO shall release the Launch Safety Agreement that provides the following information:

-   description of the WP systems/processes;
-   preliminary identification of the hazards and safety risks;
-   Identification of the CERN safety rules and host state regulations applicable to the systems/processes;
-   tailored safety advice on hazard control measures;
-   list of safety checks (including safety checks required to grant safety clearance) on the relevant systems/processes that shall be carried out by the HSE Unit during the WP lifecycle;
-   minimal contents of the WP safety file needed to meet the safety requirements.

The Launch Safety Agreement may be reviewed and updated during the different phases of the systems/sub-system's life-cycle. Any changes to the safety requirements will be subjected to change control for integration, as required, in the baseline requirements.

17.2.3    Safety folder, safety files and safety documentation

The HL-LHC project will be documented as indicated in the quality management procedure 'Safety Documentation Management' [13]. In this framework a safety folder for the HL-LHC project shall be established and maintained during the project life cycle.

-   The Work Package Engineers (WPEs), based on the applicable safety requirements, shall provide the safety documents necessary to demonstrate compliance of the systems/processes (or sub-systems/sub-processes) for which they are responsible.

- The WPE provides a description for systems and processes under their responsibility, with focus on the safety aspects. This forms the basis for the descriptive part of the WP safety file
- The WPL shall establish and maintain the WP safety file. The WP safety file shall serve as an umbrella document where the relevant safety documents, provided by the WPE, are referred to;
- The PSO shall establish and maintain the HL-LHC safety folder. The HL-LHC safety folder shall refer to the safety files of the HL-LHC work packages.

The HL-LHC safety folder will consist of the collection of approved safety files of the HL-LHC WPs. The PSO shall ensure that the HL-LHC safety folder is kept up-to-date and available to the HSE Unit and host CERN safety officers.

The safety folder for the HL-LHC shall be provided to the HSE Unit for purposes of granting safety clearance. The safety folder for HL-LHC will cover the project phase, which will extend until the end of LS3. Many of the material or organizational contributions realized in the WPs will be installed in the LHC accelerator. At this stage their safety files will become part of the LHC accelerator safety folder.

Figure 17-6 shows the organization of the HL-LHC safety documentation.

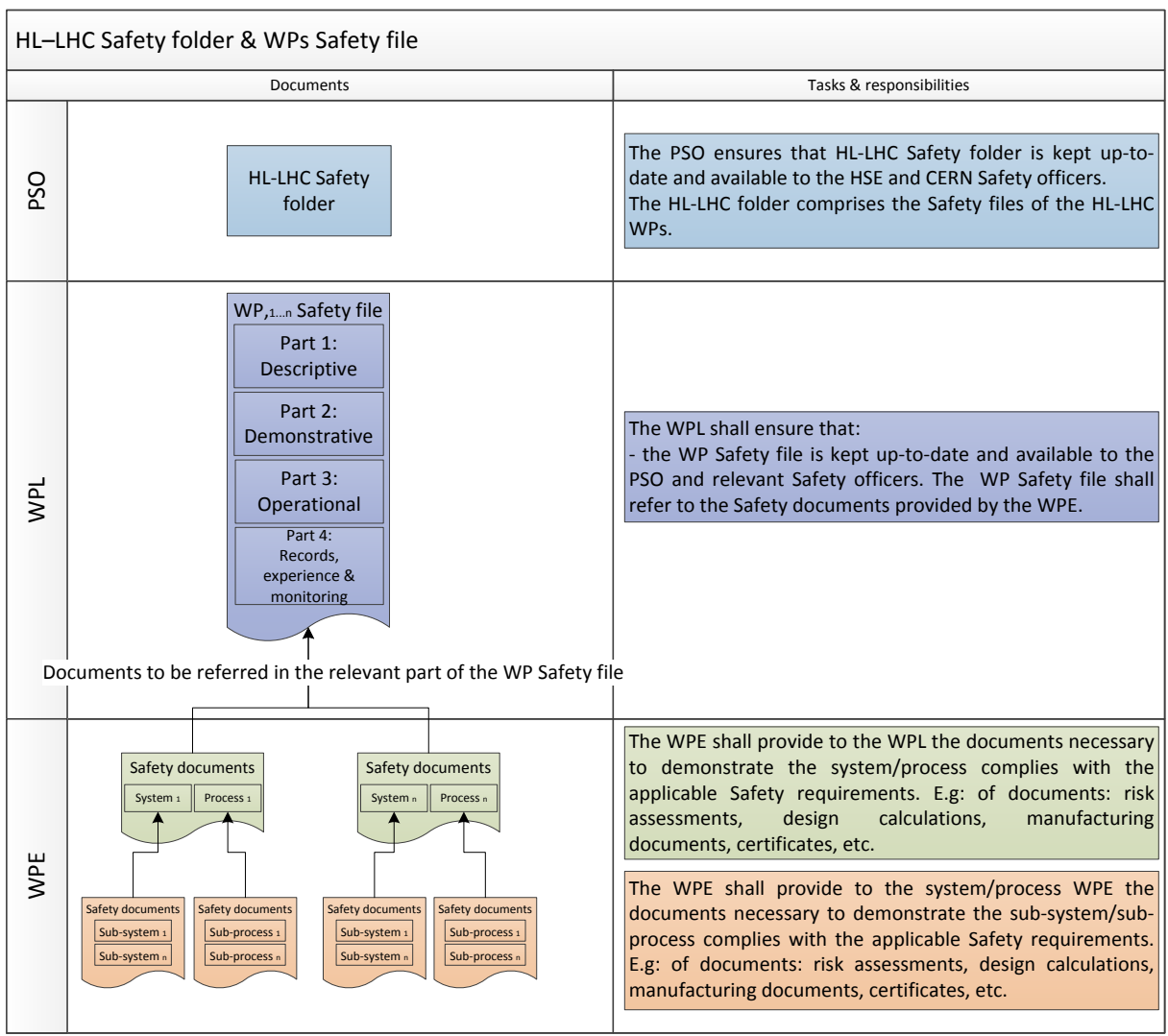

Figure 17-6: Organization of the safety documentation